\documentclass[dvipdfmx,twocolumn]{pasj01}
\begin{document} 
\Received{2018/06/14}
\Accepted{2018/10/16}%{yyyy/mm/dd}
%\Published{yyyy/mm/dd}

\title{Colors of Centaurs observed by the Subaru/Hyper Suprime-Cam and implications for their origin}

%%% begin:list of authors
% Do NOT capitalize all letters in "textsc".
\author{Haruka \textsc{Sakugawa}\altaffilmark{1,2}}
\altaffiltext{1}{Department of Planetology, Kobe University, Kobe 657-8501, Japan}
\altaffiltext{2}{Hitachi, Ltd., Tokyo 100-8280, Japan}
%\email{aaaaa@xxx.xxx.xx.xx}

\author{Tsuyoshi \textsc{Terai}\altaffilmark{3}}
\altaffiltext{3}{Subaru Telescope, National Astronomical Observatory of Japan,
National Institutes of Natural Sciences, Hilo, HI 96720, USA}
\email{tsuyoshi.terai@nao.ac.jp}

\author{Keiji \textsc{Ohtsuki}\altaffilmark{1}}
\email{ohtsuki@tiger.kobe-u.ac.jp}

\author{Fumi \textsc{Yoshida}\altaffilmark{4}}
\altaffiltext{4}{Planetary Exploration Research Center, Chiba Institute of Technology, Narashino 275-0016, Japan}
%\email{ohtsuki@tiger.kobe-u.ac.jp}

\author{Naruhisa \textsc{Takato}\altaffilmark{3}}

\author{Patryk \textsc{Sofia Lykawka}\altaffilmark{5}}
\altaffiltext{5}{School of Interdisciplinary Social and Human Sciences, Kindai University, Osaka 577-0813, Japan}

\author{Shiang-Yu \textsc{Wang}\altaffilmark{6}}
\altaffiltext{6}{Institute of Astronomy and Astrophysics, Academia Sinica, Taipei 10617, Taiwan}

%%% end:list of authors

%% `\KeyWords{}' always has to be placed before `\maketitle'.
 %Do NOT move this preamble from here!

\maketitle

\begin{abstract}
Centaurs have orbits between Jupiter and Neptune and are thought to originate from the trans-Neptunian region.
Observations of surface properties of Centaurs and comparison with those of trans-Neptunian objects (TNOs) 
would provide constraints on their origin and evolution. 
We analyzed imaging data of nine known Centaurs observed by the Hyper Suprime-Cam (HSC) installed on the Subaru Telescope with the $g$ and $i$ band filters. Using the data available in the public HSC data archive as well as those obtained by the HSC Subaru Strategic Program (HSC-SSP) by the end of June, 2017, we obtained the $g-i$ colors of the nine Centaurs.
We compared them with those of known TNOs in the HSC-SSP data obtained by \citet{te18}. 
We found that the color distribution of the nine Centaurs is similar to that of those TNOs with high orbital inclinations, but distinct from those TNOs with low orbital inclinations. 
We also examined correlations between the colors of these Centaurs and their orbital elements and absolute magnitude. 
The Centaurs' colors show a moderate positive correlation with semi-major axis, while no significant correlations between the color and other orbital elements or absolute magnitude were found for these Centaurs. 
On the other hand, recent studies on Centaurs with larger samples show interesting correlations between 
their color and absolute magnitude or orbital inclination.
We discuss how our data fit in these previous studies, and also
discuss implications of these results for their origin and evolution.
\end{abstract}

\section{Introduction}

%\noindent IMPORTANT NOTICE\\
%1. ``\verb|\draft|'' creates single column and double spaces format.\\
%2. If you comment out ``\verb|\draft|'', the output will be double column
%   and single space.\\
%3. For cross-references, the use of ``\verb|\label|, \verb|\ref|, \verb|\cite|'' 
%   and the thebibliography environment is strongly recommended. \\
%4. Do NOT use ``\verb|\def|, \verb|\renewcommand|''.\\
%5. Do NOT redefine commands provided by PASJ01.cls.\\

Recent models of planet formation show that giant planets in the solar system likely experienced
significant radial migration, causing dramatic influence on nearby small bodies \citep{ts05,wa11}.
Numerical simulations show that scattering by the migrating giant planets in the presence of the nebular gas can explain 
the degree of radial mixing in the present asteroid belt \citep{wa11}.
Migration of the giant planets after their formation and the dispersal of the nebular gas leads to
injection of a large number of small icy bodies originally in the trans-Neptunian disk
into planet-crossing orbits.
Some of these objects were presumably captured as Jupiter's Trojan asteroids \citep{mo05,ne13}, 
while others were further delivered into the asteroid belt \citep{le09} or even into the terrestrial region \citep{go05}.
The transfer of small bodies from the trans-Neptunian region to planet-crossing orbits has continued 
after planet formation, and part of them are observed today as Centaurs.
Therefore, observational studies of trans-Neptunian objects (TNOs), Centaurs and other small solar system bodies
can provide us important information about the evolution of the solar system.

Spectroscopic studies of small bodies give us detailed information about their surface properties and composition. 
However, obtaining such data for a large number of distant small bodies is a difficult and 
telescope time-consuming task.
On the other hand, photometric observation using multiple broadband filters allows us to
perform a statistical study of a large number of objects including rather faint ones (see reviews by \cite{do08} and \cite{te08}).
Using these data, we can obtain the color distribution of small body populations, which can be used to infer 
their origin and evolution.
So far, more than 2,000 TNOs have been found.
Colors of TNOs have been measured and correlations with 
orbital elements and other quantities have been extensively studied (for a review, see \cite{do08}).
These studies show that a group of TNOs with low orbital eccentricity/inclination (often called
cold classical Kuiper-belt objects) tend to be significantly redder than other TNOs (e.g., \cite{te98,te16}),
while those in other groups seem to have similar color distributions.
The size distribution of the cold classical TNOs is also distinct from other TNOs \citep{fr14}.
These pieces of observational evidence suggest that the cold TNOs likely formed approximately
where they are today, in contrast to those TNOs in other dynamical groups, which are
thought to have experienced significant scattering and radial mixing due to perturbations by the giant planets.
Also, it has been shown that small TNOs have a bimodal color distribution \citep{pe12,pe15,te16,wo17},
which are often called gray and red groups (These are called red and very red color groups in \cite{wo17}).
On the other hand, Jupiter's Trojan asteroids is a group of small bodies with the largest number of 
confirmed members ($> 7,000$) in the outer solar system. 
The overall color distribution of Jupiter Trojans is
roughly similar to the gray group in TNOs, but it shows bimodality for sub-groups 
within the Trojan population \citep{em11,wo14}.
\citet{wo16} proposes a model for the origin and evolution of the two Trojan color sub-populations 
based on the depletion or retention of H${}_2$S ice on their surfaces.

Centaurs are small solar system bodies located between the orbits
of Jupiter and Neptune, and more than 400 bodies with confirmed orbits are listed in the JPL database.
It should be noted that there are some variations in the definition of Centaurs. 
For example, the JPL database defines Centaurs as objects with $5.5  \mbox{ au} < a < 30.1 \mbox{ au}$ 
($a$ is the object's semi-major axis) with the outer boundary corresponding to Neptune's semi-major axis 
$a_{\rm N}$, while $ q = 5.2$ au ($q$ is the perihelion distance)  is used for the 
inner boundary in the Minor Planet Center database.  
Furthermore, \citet{gl08} proposes to
adopt $T_{\rm J} = 3.05$ ($T_{\rm J}$ being the Tisserand parameter with respect to Jupiter) and $q = 7.35$ au 
as the inner boundary, i.e., the boundary between Centaurs and Jupiter-family comets.
Centaurs are likely delivered from the trans-Neptunian disk by planetary perturbations \citep{le97},
but their origin is not well understood.
Owing to perturbations by the giant planets, Centaurs' dynamical lifetime is rather short ($\sim 10^6$ years;
\cite{ti03,ho04}).
Some Centaurs may evolve to become Jupiter-family comets; but
eventually, these objects will be ejected from the solar system or collide with one of the planets
 \citep{ti03}.
Therefore, studies on the origin of Centaurs allow us to better understand the process of
delivery of small icy bodies from the trans-Neptunian region as well as the origin of Jupiter-family comets.

Survey observations estimate that there are about $10^7$ Centaurs with diameter $D > 2$ km and 100 with $D > 100$ km \citep{sh00}.
The total mass of Centaurs is estimated to be about $10^{-4} M_{\oplus}$, which is as large as 
one tenth of the total mass of the main-belt asteroids.
Spectral observations have revealed that some Centaurs show signatures of various ices \citep{ba08},
but most of the information on their surface properties comes from photometric observations \citep{te08,te16,ha02,ha12}.
A notable characteristic of Centaurs' color distribution is that it is bimodal with gray and red groups
\citep{pe03,te03b}.
Recent studies show that the bimodal color distribution is common to both Centaurs and small TNOs \citep{pe12,pe15,te16,wo17}.
However, the available color data for Centaurs is still limited at present.
So, for an improved statistical study of color distribution and correlations between colors and other quantities 
(e.g.,  orbital elements), increasing the number of data with good quality is essential.
Also, a comparison of colors using a uniform set of data obtained by the same instrument for
objects in different dynamical groups can improve the quality of the comparison.

In the present work, we obtain the color distribution of nine Centaurs observed by
the Hyper Suprime-Cam (HSC) installed on the Subaru Telescope.
We examine correlations between colors and other quantities, i.e., orbital elements and absolute magnitude.
We compare the obtained color distribution with that of known TNOs obtained by HSC \citep{te18}.
The HSC data used in this work and that of \citet{te18} are of optimal quality and obtained by the same survey/instrument.
On the other hand, recent studies on Centaurs with larger samples demonstrate interesting
correlations between their color and other quantities, such as absolute magnitude \citep{pe12,te16}, and we 
will discuss how our data fit in these recent studies.
Based on these results, we discuss implications for the origin and evolution of Centaurs.
We describe the data used in the present work in Section 2.
The results of our data analysis are presented in Section 3, and we discuss and summarize our results in Section 4.

\section{Data}

\subsection{Observation and object sample}

We use imaging data obtained by the HSC. 
The HSC is a prime focus camera for the 8.2m Subaru Telescope, and consists of 116 2048 $\times$ 4096 pixel CCDs 
(104 for science, 8 for focus monitoring, and 4 for auto guiding).
It has a field of view with $1.5$ degrees in diameter with a pixel scale of 0.168 arcseconds \citep{mi12}. 
The data we used in the present work are those obtained through the HSC Subaru Strategic Program (HSC-SSP) 
by the end of June 2017 as well as those available in the public HSC data archive, which were obtained by the end of March 2016. 
The HSC-SSP project is a multi-band imaging survey with $g$, $r$, $i$, $z$, $Y$ broad-band and four narrow-band filters, 
covering $\sim 1400$ deg${}^2$ of the sky for 300 nights over $5-6$ years from March 2014 \citep{ai18}. 
As we describe below, we use data taken with the $g$ and $i$ band filters.

In the present work, we perform photometric investigations of known Centaurs using the above HSC data.
The daily ephemeris of each of known Centaurs was retrieved from the Minor Planet \& Comet 
Ephemeris Service website managed by the Minor Planet Center. We searched for known Centaurs 
with coordinates located within the area of the HSC data mentioned above at their acquisition dates, 
and checked if there was a detected source corresponding to each of those objects in the 
source catalogs using the sub-hourly ephemeris. 
The identified source was checked by visual inspection. 
If the source was judged not to be an actual
Centaur (e.g., a star, galaxy, artifact) or its image had any problems
(e.g., located at the edge of the image or too close to a very bright object), 
it was excluded from our analysis. 
We selected objects observed 
both in the $g$ and $i$ bands as the targets of this study, since previous works
show that the $g-i$ color is a useful indicator to distinguish populations of small bodies
in the outer solar system \citep{wo16,wo17,te18}. 
Based on the above criteria we finally found nine Centaurs suitable for the present study. 
Table 1 is the list of these objects and their acquisition dates in each band. 
For those objects observed multiple times with the same filter during one night,
the average time interval for observations was about 11 minutes.

\begin{figure*}
\begin{center}
 \includegraphics[width=7cm]{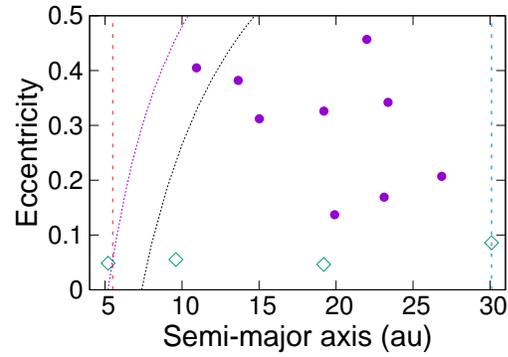} 
\caption{Distribution of the semi-major axes and eccentricities of the nine Centaurs examined in the present work (circles).
Diamonds represent the four giant planets for comparison.
The two vertical dashed lines represent $a=5.5$ au and $a=a_N$ 
($a_N$ is Neptune's semi-major axis), respectively.
The two dotted curves represent $q=5.2$ au and $q=7.35$ au, respectively.} 
\label{fig:a-e}
\end{center}
\end{figure*}

Figure \ref{fig:a-e} shows the semi-major axes and eccentricities of the above nine Centaurs (see also Table 2).
The vertical straight line at $a = 30.1$ au shows the outer boundary commonly used in various definitions of
Centaurs (Section 1).
We show three  examples of the inner boundary in Figure \ref{fig:a-e}: $a = 5.5$ au (JPL database), $q = 5.2$ au (MPC), and $q = 7.35$ au \citep{gl08}.
Most of the objects are not affected by the choice of the inner boundary.
However, the one near the inner boundary (2014 AT${}_{28}$) is included as a Centaur in JPL's and MPC's definitions, 
while it is excluded in the definition of \citet{gl08}.
It should also be noted that this object is on a retrograde orbit about the Sun (Table 2), thus it likely experienced
dynamical evolution that is quite different from other objects \citep{br12,vo13}.
Therefore, although we will measure its color together with other eight Centaurs, we will exclude this object
when we perform statistical analysis for correlations of the obtained color distribution with orbital elements and absolute magnitude (Section 3.2).

\subsection{Data reduction and color measurement}

The procedures for data reduction and analysis adopted in the present work are similar to 
those in \citet{te18}.
The data were processed with \verb+hscPipe+, which is the HSC data reduction/analysis pipeline 
developed by the HSC collaboration team \citep{bo18} based on the Large Synoptic Survey Telescope (LSST) pipeline software 
\citep{iv08,ax10,ju15}. 
First, a raw image is reduced by CCD-by-CCD procedures including bias subtraction, 
linearity correction, flat fielding, artifacts masking, and background subtraction. 
Next, the pipeline detects sources and determines the World Coordinate System (WCS) 
and zero-point magnitude of the corrected data by matching to the Pan-STARRS 1 (PS1) $3 \pi$ catalog 
\citep{to12,sc12,ma13}. 
Then, centroids, shapes, and fluxes of the detected sources are measured with several different algorithms. 
We use the sinc aperture flux \citep{bi13} with 12 pixel (i.e., $\sim 2.0$ arcseconds) radius aperture 
for photometry of the detected objects, which is also used in the estimate of the zero-point magnitude. 
Note that the zero points are translated from PS1 into the native HSC system by a color term 
\citep{ka18}. Finally, the pipeline generates source catalogs of measured values 
and flags for detected objects in each CCD.

The apparent magnitude ($m$) of the detected objects in each band is obtained based on 
the photometric zero point determined by the \verb+hscPipe+ processing mentioned above,
and is converted into the absolute magnitude ($H$) as

\begin{equation}
H=m-5\log_{10} (R\cdot \Delta)-\beta \cdot \alpha,
\end{equation}
where $R$ and $\Delta$ are the heliocentric and geocentric distances in units of au, respectively;
$\alpha$ is the solar phase angle; and $\beta$ is the phase coefficient. 
Following \citet{te18}, we assume a constant value of 
$\beta=0.11$ mag deg${}^{-1}$ \citep{al16} for both $g$ and $i$ bands.

For each object, we calculated a weighted mean of the normalized flux ($F_x$) in a band $x$ ($x$ is 
either $g$ or $i$) and its uncertainty ($\sigma_{F_x}$) with the following equations:

\begin{equation}
F_x=\frac{\sum_{j=1}^{N_x} \sigma^{-2}_{x,j} F_{x,j} }{ \sum_{j=1}^{N_x} \sigma^{-2}_{x,j} },
\end{equation}
\begin{equation}
\sigma_{F_x} = \sqrt{ \frac{ 1 }{ \sum_{j=1}^{N_x} \sigma^{-2}_{x,j} } 
                    + \frac{ \sum_{j=1}^{N_x} \sigma^{-2}_{x,j} (F_{x,j}-F_x)^2 }
                           { (N_x-1) \sum_{j=1}^{N_x} \sigma^{-2}_{x,j} } },
\end{equation}
where $F_{x,j}$ and $\sigma_{x,j}$ are the normalized $x$-band flux given from the measured absolute 
magnitude at each epoch ($j$ = 1, 2, $\cdots$, $N_x$; $N_x$ is the number of $x$-band data) and its 
uncertainty, respectively.
The average absolute magnitude and its uncertainty, $H_x$ and $\sigma_{H_x}$, are converted from 
$F_x$ and $\sigma_{F_x}$, respectively.
Then, the $g-i$ color is derived from $H_g -H_i$.

In addition to the photometric error, brightness variation due to the small body's rotation also 
contributes to uncertainty in the color measurement. 
According to the analysis by \citet{du09}, the mean rotation period and lightcurve amplitude
of Centaurs are 6.75~hr and 0.26~mag, respectively. 
Using a Monte Carlo method, we generated synthetic lightcurves assuming a sinusoidal brightness 
fluctuation with 6.75~hr period, 0.26~mag amplitude, and random initial phase angles at the actual 
acquisition epochs of each object/band.
The error in color caused by rotation is evaluated by the standard deviation ($\sigma_{\rm rot}$) 
of the color offsets.
The typical value is $\sigma_{\rm rot} \sim 0.1$~mag.
Finally, the total color uncertainty is estimated from 
$\sqrt{ \sigma_{H_g}^2 + \sigma_{H_i}^2 + \sigma_{\rm rot}^2 }$.

The averaged absolute magnitudes ($H_g$ and $H_i$) and $g-i$ colors obtained for all the sample 
objects are shown in Table~2.
Note that the magnitudes and colors are expressed in the Vega magnitude system hereafter.

\section{Results}

\subsection{Color distribution}

\begin{figure*}
\begin{center}
\includegraphics[width=13cm]{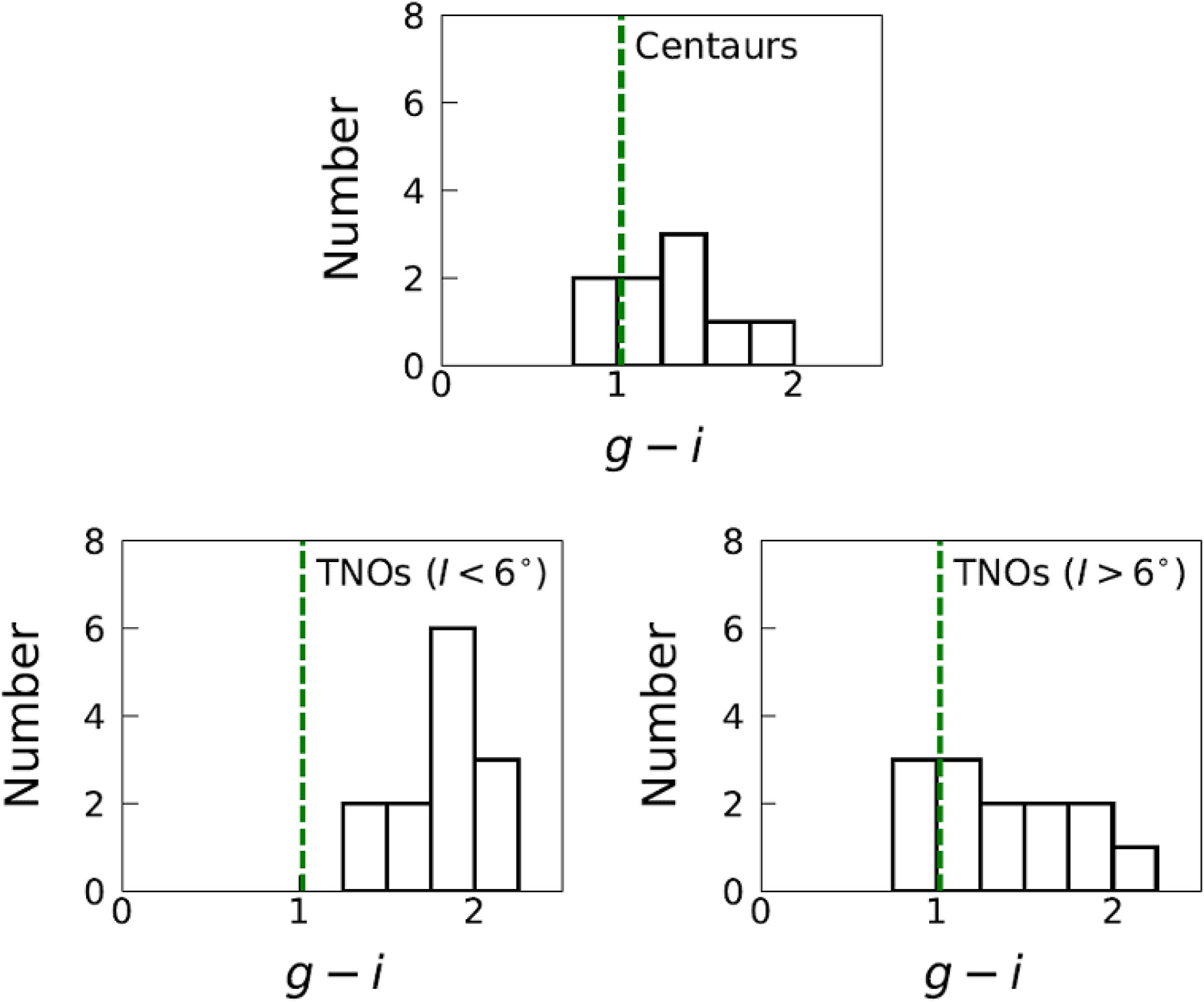}
\end{center}
\caption{Top panel: Distribution of $g-i$ color of the nine Centaurs examined in the present work.
The vertical dashed line represents the solar color ($g-i = 1.02$).
Bottom panels: Distributions of $g-i$ color of TNOs 
examined by \citet{te18} based on the HSC-SSP data.
The left panel shows the distribution of 13 objects with low-inclination ($I < 6^{\circ}$),
and the right panel shows that of 13 objects with high-inclination ($I > 6^{\circ}$).
Data for the bottom panels were taken from \citet{te18}.} 
\label{fig:colordist}
\end{figure*}

The color distribution of the nine Centaurs obtained in the present work is shown in the 
top panel of  Figure \ref{fig:colordist}.
The solar color is given by \citet{ho06} and is converted into the HSC band system with color term correction (see Table 3 in \cite{te18}).
Here, we use $g-i = 1.02$ mag, which is shown with the dashed line for comparison in Figure \ref{fig:colordist}.
We find that the colors of the nine Centaurs are distributed over the range from neutral to slightly red colors.
The $g-i$ color of Chiron obtained by the present work was about 0.96 mag, 
which is consistent with previous works \citep{ha90,me90,lu90,ha02,ro03}.
We do not see clear bimodality in the color distribution that was found in previous works 
(\cite{te03a,pe03,te03b}; see a review by \cite{te08}), 
owing to the small number of objects examined in the present work; 
comparison with previous works with larger samples will be discussed in Section 3.2.

The bottom panels of Figure \ref{fig:colordist} show similar plots for TNOs obtained by \citet{te18}.
\citet{te18} measured colors of 30 known TNOs also using HSC-SSP data, and examined correlations 
between colors and other quantities, such as orbital elements.
They found that dynamically hot classical TNOs and scattered objects have similar color distributions, while
dynamically cold classical TNOs are distinctly redder in the $g$ and $r$ bands.
The left bottom panel of Figure \ref{fig:colordist} shows the distribution for 13 objects 
with low orbital inclinations ($<6^{\circ}$) , while
the right panel shows the case of 13 objects with high orbital inclinations ($> 6^{\circ}$)
(Note that among the 30 objects examined by \citet{te18}, resonant objects are excluded in these plots,
because their orbital evolution was likely different from other TNOs).
By comparing the color distributions in Figure \ref{fig:colordist}, 
we found that the distribution of the Centaurs is rather similar to that of TNOs with high orbital inclinations. 
In particular, both populations have a peak at around  $1.0 \lesssim g-i \lesssim1.5$, and many objects have colors similar to 
the solar color.

Next, we examined the statistical significance of the two populations of 
TNOs (i.e., high-$I$ and low-$I$ populations) as the dynamical  source of Centaurs
by performing the Kolmogorov-Smirnov (K-S) test \citep{pr92}.
The K-S test is a statistical measure indicating whether two data sets are drawn from a single distribution function.
We found that the probability that the color distributions of our Centaur sample and
Terai et al.'s low-$I$ TNOs are the same by chance is only 0.49\%, which means that the two distributions are significantly different.
We found a similar result (0.38\%) after using a variant test appropriate for a small number of samples 
\citep{mi56}.
On the other hand, our K-S test \citep{pr92} shows 88\% probability that the color distributions 
are the same for our Centaur sample and Terai et al.'s high-$I$ TNOs.

These results confirm the trends seen in Figure \ref{fig:colordist}  that the color distribution of the Centaurs is similar to that of 
those TNOs with high inclinations.
This fact implies both populations have similar surface properties, which suggests that 
members of these populations formed at similar radial locations in the solar system.
On the other hand, the color distribution of the Centaurs is distinct from that of TNOs with 
low inclinations, which have many more redder objects.
These results support the idea that high-$I$ TNOs are the main source of currently observed Centaurs.
This assumes that the colors of Centaurs are not significantly affected during their dynamical evolution.
This view is also consistent with recent works with bigger samples that show that the bimodal color distribution
is seen not only for Centaurs but also both small and large TNOs \citep{pe12,fr12,te16,wo17}.

\subsection{Correlations between color and other quantities}

We investigated correlations of the obtained colors of the nine Centaurs (including the retrograde object  2014AT${}_{28}$) with 
other quantities, i.e., their orbital elements and absolute magnitude.
If there are any correlations between colors and current orbital elements, it may suggest that
the Centaurs acquired their colors through some processes after they were injected into their current orbits.
For example, an object's  semi-major axis is related to its mean temperature determined
by the distance from the Sun, while its perihelion distance reflects the maximum
temperature the body experiences on its current orbit.
The orbital eccentricity and inclination of an object may provide some information about
its impact velocity at its current orbit with other bodies  \citep{ha12}.
On the other hand, dynamical modeling shows that it is difficult to infer an object's collisional history
from its current orbital properties alone, because its collisional evolution also depends on
dynamical properties of impactors \citep{th03}.
Furthermore, if the process that determined the object's color is size-dependent, 
some correlation between color and absolute magnitude would be expected.

\begin{figure*}
\begin{center}
\includegraphics[width=13cm]{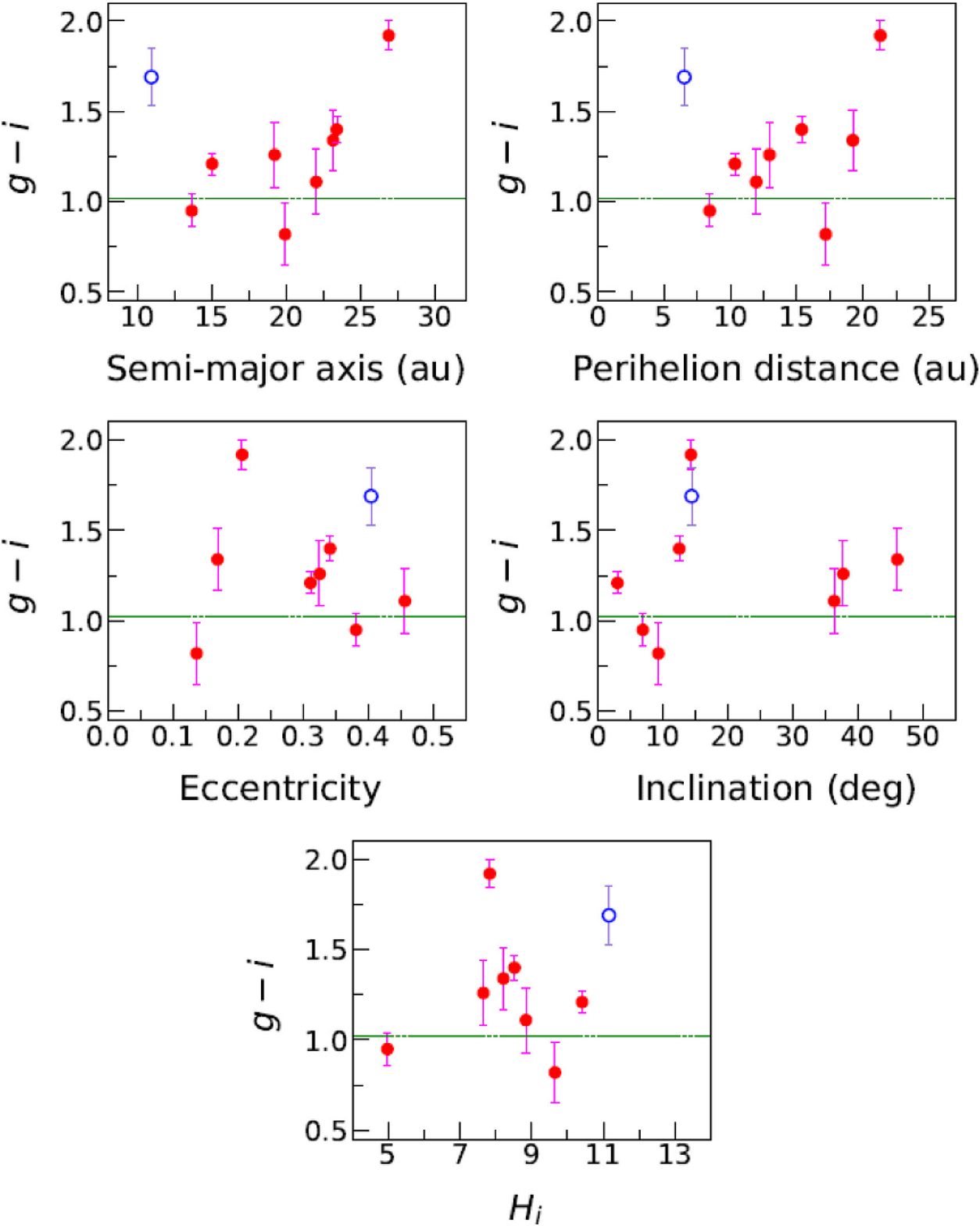}
\end{center}
\caption{Dependence of $g-i$ color on semi-major axis ($a$), perihelion distance ($q$), 
eccentricity ($e$), inclination ($I$),  and measured absolute magnitude in the $i$ band ($H_i$) 
for the nine Centaurs examined in the present work.
The open circle represents 2014 AT${}_{28}$, which has a retrograde orbit and is not regarded as a 
Centaur in the definition by \citet{gl08}(see Figure 1) . 
The dashed line represents the solar color.}
\label{fig:colorcorr}
\end{figure*}

The top and middle panels of Figure \ref{fig:colorcorr} show relationships between the colors and orbital elements 
for the nine Centaurs.
The dashed line represents the solar color.
The open circles denotes 2014 AT${}_{28}$; since this object has a retrograde orbit with
$I \simeq 166^{\circ}$, the value of $180^{\circ} - I$ is shown for this object, instead of $I$.
Also, the relationship between color and absolute magnitude is shown in the bottom panel
of the same figure.
We notice that the retrograde object  2014AT${}_{28}$ shows distinct behaviors 
in the plots of semi-major axis and perihelion distance,  which may reflect its orbital evolution
different from others \citep{br12,vo13}.
If we exclude this object, colors of other Centaurs seem to have positive correlation with
their semi-major axis and perihelion distance.
However,  the correlation is not clear, presumably because
of the small number of data points.

We examined statistical significance of the correlations between
colors and orbital elements as well as absolute magnitude for the eight Centaurs on prograde orbits, 
using Spearman rank-order correlation coefficient $r_s$.
For $N$ pairs of variables ($x_i$, $y_i$),  $r_s$ is given as 
\begin{equation}
 r_s=1-\frac{6\sum_{i=1}^N(X_j-Y_j)^2}{N^3-N},
\end{equation}
where $X_i$ and $Y_i$ are the ranks of $x_i$ and $y_i$, respectively.
$r_s$ can be converted into Student's $t$ statistic with the degree of freedom 
$\nu = N - 2$ by

\begin{equation}
 t=r_s\sqrt{\frac{N-2}{1-{r_s}^2}}.
\end{equation}
The probability ($P_t$) for the null hypothesis that the two distributions have no correlation
can be  calculated by the incomplete beta function
 \begin{equation}
 I_x(a,b)=\frac{\Gamma(a+b)}{\Gamma(a)\Gamma(b)}\int_0^x t^{a-1}(1-t)^{b-1}dt,
 \end{equation}
where $\Gamma$ is the gamma function.
Also, $x = \nu/(\nu + t^2)$, $a = \nu/2$, and $b = 1/2$.

The results of the above statistical analysis are shown in Table 3.
We found that the $g-i$ color has a potentially significant correlation  
(i.e., the null hypothesis is rejected at 0.05 level of significance) 
with semi-major axis.
We found a similar result when we perform the test using a table for the case of small numbers \citep{ra89}.
Such a correlation was not found in the analysis of \citet{te18} for TNOs 
observed by HSC-SSP.
Although this positive correlation
may reflect a certain evolutionary process of the Centaurs examined in the present work,
further investigation is desirable to provide definite conclusions.
On the other hand, Table 3 shows that the correlation between color and 
perihelion distance is not significant for the objects we examined.

Previous works show that the onset of cometary activity in Centaurs and the 
disappearance  of the very red material on their surfaces seem to begin at about $q \sim 10$ au \citep{je09,je15}.
\citet{je09} argued that crystallization of amorphous ice, with the concomitant release of trapped volatiles, 
is the leading candidate process for the explanation of these observations. 
The absence of cometary activity in the twenty Centaurs discovered by \citet{ca18} is consistent with this scenario.
In fact, the top-right panel of Figure \ref{fig:colorcorr} shows no
object redder than the solar color for $q \lesssim 10$ au, which seems consistent with the above argument.
However, it is difficult to confirm the critical radial location of the change of the colors from our results 
since there is only one object (Chiron) on prograde orbits with $q < 10$ au in our sample.
Also, the color distribution of inactive Centaurs presented in \citet{je15} seems to be rather independent of $q$
at $q \gtrsim 10$ au.
Further investigation using a larger number of Centaurs including those with smaller perihelion distance is desirable.

In addition, we found no correlations of the $g-i$ color with other orbital elements. 
By analyzing the colors of TNOs observed by the HSC-SSP,
\citet{te18}  found that the color distribution as a function of orbital eccentricity can be divided
into two groups, i.e., one group with low eccentricity and red color, and another group
with relatively high eccentricity and neutral color (see their Figure 11).
They also found that colors of those objects in the latter group have a positive 
correlation with eccentricity, while they show a negative correlation 
with inclination.
Such trends are not found
for the eight Centaurs examined in the present work, which was also based on the HSC data.
Recently, from the analysis of color date of 61 Centaurs, \citet{te16} found that red Centaurs have 
significantly smaller orbital inclinations than gray Centaurs, while correlations between color 
and other orbital elements were not found (see also \cite{te08}).
Further studies are required to understand implications of these observations for the 
origin and evolution of Centaurs.

As for the dependence of colors on absolute magnitude, \citet{ha12} showed that
color distribution of small bodies in the outer solar system seems to be size-independent, 
except that large ones are slightly bluer than the smaller ones in the near infrared 
(i.e., in the $J-H$ color).
On the other hand, \citet{pe12} found that the bimodal color distribution of Centaurs and TNOs 
is size-dependent (see also \cite{pe15}).
They found that not only Centaurs but also small TNOs with $H_R \gtrsim 6.8$ 
($H_R$ is the $R$-band absolute magnitude) have bimodal color distribution 
(see also \cite{fr12,wo17}), while such a bimodality is not seen for larger TNOs with  $5 \lesssim H_R \lesssim 6.8$.
\citet{pe12} noted that still larger TNOs with $H_R \lesssim 5$ show another bimodal color distribution. 
More recently, \citet{te16} found that their entire sample of TNOs and Centaurs exhibits bimodal colors.
They also found that the red Centaurs have average albedos about a factor of two larger than the gray Centaurs,
while both populations have broadly similar absolute magnitude distribution with a slight possibility of 
fainter absolute magnitude values for the red ones.
This suggests that red Centaurs may have smaller diameters than gray Centaurs.
The tendency that red objects have larger albedos has also been reported for Jupiter's Trojans 
(\cite{wo14}; see also \cite{em11} and \cite{gr12}).
Although we did not find any correlation between the $g-i$ color and absolute magnitude
for the nine Centaurs we examined, the above studies suggest that examination of size-dependence in larger sample
including bodies in other dynamical groups is important for a better understanding of the meaning of the color distribution of these bodies.

In order to understand how our data for the nine Centaurs fit in these previous studies with larger samples, 
we calculated spectral slopes from the $g-i$ color for the nine objects in our sample and from the $B-R$ color for the 61 
Centaurs obtained by \citet{te08} and \citet{te16}, and compared them in Figure \ref{fig:slope}
(see Appendix for the calculations of these spectral slopes).
Although we did not see bimodal color distribution for our nine objects alone (Figure \ref{fig:colordist}),
in Figure \ref{fig:slope} seven objects of our sample seem to be in the gray group and 
the other two seem to belong to the red group in the bimodal distribution found by \citet{te08} and \citet{te16}.
This strengthens  the importance of using a large sample to find
correlations that may yield useful constraints on the origin and
evolution of these small bodies \citep{pe15,te16}.

\section{Conclusions and Discussion}

In the present work, we examined color distribution of nine Centaurs observed by the Hyper Suprime-Cam (HSC)
installed on the Subaru Telescope, and compare the results with the color distribution of
TNOs also obtained by HSC \citep{te18}.
We found that the color distribution of the Centaurs we observed is not significantly different from that of
the TNOs with high orbital inclinations ($ > 6^{\circ}$),
while it is distinct from that of cold classical TNOs with low orbital inclinations ($ < 6^{\circ}$).
This result suggests that TNOs with high orbital inclinations and Centaurs have a common origin.
This view is also consistent with recent works using larger samples that show that the bimodal color distribution
is seen not only for Centaurs but also both small and large TNOs \citep{pe12,fr12,te16,wo17}.
Although we did not see the bimodal color distribution found in previous studies \citep{pe03,te03b}
in our sample alone, by comparing spectral slopes obtained from
our data with those calculated from $B-R$ colors of larger samples obtained by previous works \citep{te08,te16},
we found that seven and two Centaurs in our sample belong to the gray and red groups, respectively.

We also investigated correlations of the color distribution of these Centaurs with  other 
quantities, i.e, orbital elements and absolute magnitude.
While we found a potentially significant positive correlation between the color and semi-major axis,
no significant correlations were found with other orbital elements.
On the other hand, recent works with larger samples show that red Centaurs have significantly smaller
orbital inclinations than gray Centaurs, while no significant correlations were found 
between the color and other orbital elements \citep{te16}.
We did not find correlation between the color and absolute magnitude for the nine objects we examined, 
but recent works with larger samples show that the bimodal color distribution is size-dependent \citep{pe12,fr12,pe15,te16,wo17}.
Further investigation with a larger number of samples is desirable to confirm these findings.

The observed color distribution of Centaurs may be linked to space weathering, collisional evolution,
cometary activity \citep{ha02,je09,je15}, or a combination of these and other processes
(e.g., intrinsic composition for distinct formation birthplaces).
Also, Centaurs can possess ring systems \citep{br14,ru15,or15}.
Models for the formation of such ring systems include
collisional ejection from the parent body's surface, disruption of primordial satellite,
dusty outgassing \citep{pa16}, and partial disruption of the parent body during a planetary encounter \citep{hy16}.
The possession of ring systems \citep{or15} as well as their formation process may also be linked to the Centaur color bimodality.
Further observations of Centaurs, TNOs, and other small solar system bodies are essential to
better understand the origin and evolution of Centaurs and their parent population.

\vspace{4ex}
\noindent
{\bf \textsf{Funding}}

\noindent
{\scriptsize
This work was supported by JSPS KAKENHI Nos. 18K13607 (T. T.), 15H03716, 16H04041 (K. O.),
and 16K05546 (F. Y.). 
}

\begin{ack}
We thank the anonymous reviewer for detailed comments and suggestions, which greatly improved the 
presentation of the manuscript. 
We also thank Fumihiko Usui for discussion and comments on our work, and Chien-Hsiu Lee for comments on the manuscript.
This study is based on data collected at Subaru Telescope and retrieved from the HSC data archive system, which are operated by Subaru Telescope and Astronomy Data Center, National Astronomical Observatory of Japan. The Hyper Suprime-Cam (HSC) collaboration includes the astronomical communities of Japan and Taiwan, and Princeton University. The HSC instrumentation and software were developed by the National Astronomical Observatory of Japan (NAOJ), the Kavli Institute for the Physics and Mathematics of the Universe (Kavli IPMU), the University of Tokyo, the High Energy Accelerator Research Organization (KEK), the Academia Sinica Institute for Astronomy and Astrophysics in Taiwan (ASIAA), and Princeton University. Funding was contributed by the FIRST program from the Japanese Cabinet Office, the Ministry of Education, Culture, Sports, Science and Technology (MEXT), the Japan Society for the Promotion of Science (JSPS), the Japan Science and Technology Agency (JST), the Toray Science Foundation, NAOJ, Kavli IPMU, KEK, ASIAA, and Princeton University. This paper makes use of software developed for the Large Synoptic Survey Telescope. We thank the LSST Project for making their code available as free software at http://dm.lsst.org. The Pan-STARRS1 Surveys (PS1) have been made possible through contributions of the Institute for Astronomy, the University of Hawaii, the Pan-STARRS Project Office, the Max-Planck Society and its participating institutes, the Max Planck Institute for Astronomy, Heidelberg and the Max Planck Institute for Extraterrestrial Physics, Garching, the Johns Hopkins University, Durham University, the University of Edinburgh, Queen's University Belfast, the Harvard-Smithsonian Center for Astrophysics, the Las Cumbres Observatory Global Telescope Network Incorporated, the National Central University of Taiwan, the Space Telescope Science Institute, the National Aeronautics and Space Administration under Grant No. NNX08AR22G issued through the Planetary Science Division of the NASA Science Mission Directorate, the National Science Foundation under Grant No. AST-1238877, the University of Maryland, Eotvos Lorand University (ELTE), and the Los Alamos National Laboratory. 
\end{ack}

\appendix 
\section*{
Calculation of spectral slopes from $B-R$ and $g-i$ colors
}

Let $m_{F}$ and $m_{V}$ be the magnitudes of an object in
filter $F$ and in the $V$ band, respectively, and  $m_{F, {\rm sun}}$ and $m_{V, {\rm sun}}$
be those for the Sun.
Then, the reflectance of the object normalized to unity at the $V$ central wavelength 
is given by

\begin{equation}
R_{F,V} = 10^{-0.4[(m{F}-m{V})-(m{F, {\rm sun}}-m{V, {\rm sun}})]} = R_{F}/R_{V},
\end{equation}

\noindent
where $R_{F}$ and $R_{V}$ are the reflectance at each band.
When the reflectance of an object is measured in two filters F${}_1$ and  F${}_2$,
the spectral slope (or spectral gradient) is given as (e.g., \cite{do08})

\begin{equation}
S = (R_{F2,V}-R_{F1,V})/(\lambda_2-\lambda_1)	
\end{equation}

\noindent
where $\lambda_{F}$ represents the central wavelength of filter $F$.
Alternatively, when the reflectance is normalized at  $\lambda_1$,
the slope is given by

\begin{equation}
S = (R_{F2,F1}-1)/(\lambda_2-\lambda_1)
\end{equation}

\noindent
where $R_{F1,F2} = R_{F2}/R_{F1}$.

Here, we assume that colors of an object are obtained through observations
using two pairs of filters, i.e., one observation with $B$ and $R$ filters and another 
with $g$ and $i$ filters. 
We will derive an expression for the spectral slope from the $B-R$ color 
normalized at the central wavelength of $g$ filter, $\lambda_g$, so that
it can be directly compared with the slope obtained from the $g-i$ color of the
same object.
The spectral slope obtained from $g-i$ color normalized at $\lambda_g$ is
given by  \citep{wo17}

\begin{equation}
S_{g-i, g}	=  \left\{10^{0.4[(mg-mi)-(mg{\rm ,sun} - mi{\rm ,sun})]} - 1\right\}/ (\lambda_i-\lambda_g)
\label{eq:sgi-g}
\end{equation}

\noindent
On the other hand, let the spectral slopes obtained from the $B-R$ color 
normalized at $\lambda_g$ and $\lambda_{B}$ be $S_{B-R, g}$ and $S_{B-R, B}$ respectively.
Then we have the following relationship between these two slopes:

\begin{equation}
S_{B-R, g}	= f_{g-B} \times S_{B-R, B}
\label{eq:sbr-g}
\end{equation}

\noindent
where

\begin{equation}
f_{g-B} = 10^{0.4[(mg-m{B})-(mg{\rm ,sun}-m{B, {\rm sun}})]}
\end{equation}

\noindent
Here, we assume that the values of the spectral slopes normalized at 
$\lambda_{B}$ is the same for the two wavelength ranges
$\lambda_{B} \leq \lambda \leq \lambda_g$ and
$\lambda_{B} \leq \lambda \leq \lambda_{R}$, i.e.,

\begin{equation}
\begin{array}{ll}
&\left\{10^{0.4[(mB-mg)-(m{B, {\rm sun}} - mg{\rm ,sun})]} - 1\right\}/ (\lambda_g-\lambda_B)\\ 
= &\left\{10^{0.4[(m{B}-m{R})-(m{B, {\rm sun}}-m{R, {\rm sun}})]} - 1\right\}/ (\lambda_R-\lambda_B) 
\end{array}
\label{eq:assump}
\end{equation}

\noindent
The right-hand side of Eq. (\ref{eq:assump}) is equal to $S_{B-R,B}$.
Using $f_{g-B}$, we can rewrite (\ref{eq:assump}) as

\begin{equation}
	(f_{g-B}^{-1} - 1)/ (\lambda_g-\lambda_B) = S_{B-R,B}
\end{equation}

\noindent
Solving this equation for  $f_{g-B}$, we obtain

\begin{equation}
f_{g-B}= 1/ \left\{(\lambda_g-\lambda_B) S_{B-R,B} + 1\right\}
\end{equation}

\noindent
Substituting this into (\ref{eq:sbr-g}), we finally obtain

\begin{equation}
S_{B-R,g}	= S_{B-R,B} / \left\{(\lambda_g-\lambda_B)S_{B-R,B} + 1\right\}
\label{eq:sbr-final}
\end{equation}

\begin{figure*}
\begin{center}
 \includegraphics[width=7cm]{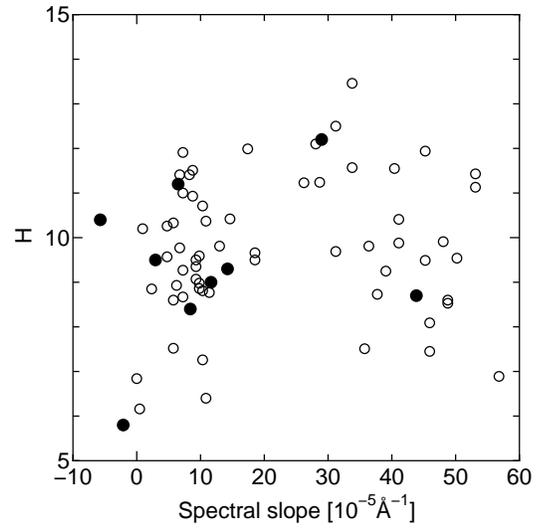} 
 \end{center}
 \caption{Relationship between absolute magnitude and spectral slope for Centaurs.
 Filled circles represent those calculated from the $g-i$ color for the nine objects 
 examined in the present work. The absolute magnitudes of these objects were obtained
 from the Minor Planet Center.
 Open circles represent those calculated from the $B-R$ color for the 61 objects examined
 by \citet{te08} and \citet{te16}.
 Methods of calculation of the spectral slopes are described in Appendix.
} 
\label{fig:slope}
\end{figure*}

\noindent
Using Eqs.(\ref{eq:sgi-g}) and (\ref{eq:sbr-final}), we can directly compare the spectral slopes 
obtained from $B-R$ and $g-i$ colors, both normalized at $\lambda_g$ (Figure \ref{fig:slope}).

%
%\section{Case of two or paragraphs}
%
%\section{Case of two or paragraphs}

%%%
% See the manual for the detail.
%%%

\newpage
\begin{table*}
\caption{Measured Centaurs and their acquisition dates for each filter${}^*$}
\begin{center}
\begin{tabular}{lcl}
 \hline
 Object & Filter & UT Date \\ 
\hline
1977 UB (Chiron) & $g$ & 2015-10-14 (1), 2016-08-28 (5) \\
          & $i$ & 2016-09-05 (1) \\
1998 QM${}_{107}$ (Pelion) & $g$ &2015-10-07 (10), 2017-01-26 (1) \\
                & $i$ & 2015-10-07 (11) \\
2006 SX${}_{368}$ & $g$ & 2014-11-20 (4) \\
                & $i$ & 2014-11-19 (4) \\
2012 DD${}_{86}$ & $g$ & 2015-03-25 (4), 2016-04-04 (6), 2016-04-05 (2), 2016-04-06 (3),\\
              &  & 2017-03-05 (1), 2017-03-29 (1) \\
              & $i$ & 2015-03-20 (4), 2016-04-02 (4), 2017-04-27 (3) \\
2013 RG${}_{98}$ & $g$ & 2015-01-20 (5) \\
              & $i$ & 2015-01-17 (5) \\
2014 AT${}_{28}$ & $g$ & 2015-10-14 (1) \\
              & $i$ & 2014-09-24 (9) \\
2014 GP${}_{53}$ & $g$ & 2015-03-25 (5), 2016-04-04 (1) \\
              & $i$ & 2014-03-28 (4), 2015-03-20 (5), 2015-03-22 (1), 2016-04-09 (2) \\
2014 SR${}_{303}$ & $g$ & 2014-07-06 (4), 2014-09-23 (2) \\
               & $i$ & 2014-07-08 (7), 2014-09-23 (1) \\
2015 DB${}_{216}$ & $g$ & 2016-01-11 (5) \\
                & $i$ & 2016-01-12 (6) \\ \hline
\end{tabular}
\end{center}
\begin{tabnote} \normalsize
${}^*$The number of data points for each date is displayed in parentheses.
\end{tabnote}
\end{table*}

\begin{table*}
\caption{Orbital elements, absolute magnitudes, and $g-i$ colors of Centaurs*}
\begin{center}
\begin{tabular}{lrrrrrrr}\hline
Object &                             $a$ (au)  & $q$ (au) & $e$   & $I$ (deg)& $H_g$ (mag)    & $H_i$ (mag)        & $g-i$ (mag) \\ \hline
1977 UB (Chiron)              & 13.648    & 8.431     & 0.382 &  6.950    & 5.92$\pm$0.01  & 4.97$\pm$0.01  & 0.95$\pm$0.09 \\
1998 M${}_{107}$ (Pelion)  & 19.911    & 17.178   & 0.137 &  9.348    & 10.47$\pm$0.10 & 9.65$\pm$0.04 & 0.82$\pm$0.17 \\  
2006 SX${}_{368}$             & 21.997    & 11.945   & 0.457 & 36.325   & 9.95$\pm$0.00  & 8.84$\pm$0.01  & 1.11$\pm$0.18 \\ 
2012 DD${}_{86}$              & 23.373    & 15.372   & 0.342 & 12.498   & 9.91$\pm$0.04  & 8.52$\pm$0.03  & 1.40$\pm$0.07 \\ 
2013 RG${}_{98}$              & 23.125    & 19.224   & 0.169 & 46.036   & 9.55$\pm$0.02  & 8.21$\pm$0.03   & 1.34$\pm$0.17 \\
2014 GP${}_{53}$              & 26.860    & 21.308    & 0.207 & 14.256  & 9.75$\pm$0.04   & 7.83$\pm$0.01  & 1.92$\pm$0.08 \\
2014 SR${}_{303}$            & 15.017    & 10.333    & 0.312 &   3.035   & 11.62$\pm$0.02 & 10.41$\pm$0.02 & 1.21$\pm$0.06 \\
2014 AT${}_{28}$              & 10.942    &   6.512    & 0.405 & 165.559 & 12.84$\pm$0.00 & 11.16$\pm$0.01 & 1.69$\pm$0.16 \\
2015 DB${}_{216}$            & 19.211    & 12.944    & 0.326 &   37.701 & 8.91$\pm$0.00   & 7.65$\pm$0.01  & 1.26$\pm$0.18\\ \hline
\end{tabular}
\end{center}
\begin{tabnote} \normalsize
*Semi-major axis ($a$), perihelion distance ($q$), eccentricity ($e$), inclination ($I$) , 
measured absolute magnitudes in the $g$ and $i$ bands ($H_g$ and $H_i$),
and the $g-i$ color. Note that 2014 AT${}_{28}$ has a retrograde orbit.
\end{tabnote}
\end{table*}%

\begin{table*}
\caption{Results of correlation test*}
\begin{center}
\begin{tabular}{cccccc}\hline
&$a$&$q$&$e$&$I$&$H_i$ \\ \hline
$r_s$&0.738&0.571&-0.095&0.429&-0.238 \\
$P_t$&0.037&0.139&0.823&0.289&0.570 \\ \hline
\end{tabular}
\end{center}
\begin{tabnote} \normalsize
*Significance of correlation between $g-i$ color and other quantities: Semi-major axis ($a$), 
perihelion distance ($q$), eccentricity ($e$), inclination ($I$) , and absolute magnitude in the $i$ band ($H_i$).
$r_s$ is the Spearman rank-order correlation coefficient, and $P_t$ is the probability that such correlation 
value would be seen when the null hypothesis of no correlation is true.
Note that the retrograde object 2014 AT${}_{28}$ was excluded from the analysis of correlations.
\end{tabnote}
\end{table*}

\newpage

\newpage

\newpage

\end{document}